\documentclass[conference]{IEEEtran}
\ifCLASSINFOpdf
\else
\fi
\usepackage{amsmath}
\usepackage{amssymb}
\usepackage{graphicx}
\usepackage{subcaption}

\usepackage{xcolor}

\usepackage[draft, authormarkup=none, todonotes={textsize=tiny, textwidth= 40pt}]{changes}
\setdeletedmarkup{} 
\definechangesauthor[name={}, color=orange]{pz}
\begin{document}
%
\title{Benchmarking Wireless Representations: High-Dimensional vs. Compressed Embeddings for Efficiency and Robustness\vspace{-2mm}}


\author{\IEEEauthorblockN{Murilo Batista\IEEEauthorrefmark{1}, Shirin Salehi\IEEEauthorrefmark{1}, Saeed Mashdour\IEEEauthorrefmark{2}, Paul Zheng\IEEEauthorrefmark{1},
Rodrigo C. de Lamare\IEEEauthorrefmark{2}\IEEEauthorrefmark{4} and Anke Schmeink\IEEEauthorrefmark{1}
}

\IEEEauthorblockA{\IEEEauthorrefmark{1} Chair of Information
Theory and Data Analytics (INDA), RWTH Aachen University, Aachen, Germany}

\IEEEauthorblockA{\IEEEauthorrefmark{2} Centre for Telecommunications Studies, Pontifical Catholic University of Rio de Janeiro, Rio de Janeiro 22541-041, Brazil}    
 
\IEEEauthorblockA{\IEEEauthorrefmark{4} School of Physics, Engineering and Technology, University of York, United Kingdom}


    murilo.batista@rwth-aachen.de, \{shirin.salehi, paul.zheng, anke.schmeink\}@inda.rwth-aachen.de,\\  
    smashdour@gmail.com, delamare@puc-rio.br \vspace{-5mm}
}

%


\maketitle

\begin{abstract}
Building on recent advances in representation learning for wireless channels, this work investigates the cost-benefit trade-offs of high-dimensional channel embeddings in practical systems. We benchmark multiple wireless representations: high-dimensional learned embeddings from a wireless foundation model, compact autoencoder-based representations with significantly lower dimensionality, and raw data baselines, evaluating their performance across diverse downstream tasks. We then systematically analyze data efficiency, noise robustness, and computational complexity, explicitly characterizing the resource overhead associated with high-dimensional embeddings.
Beyond standard tasks such as line-of-sight/non-line-of-sight (LoS/NLoS) classification and beam selection, we introduce power allocation as a new downstream task. 
Our results reveal clear trade-offs: while high-dimensional embeddings can perform well in few-shot regimes for certain tasks, they incur substantial latency and parameter overhead. In contrast, compressed latent representations learned by autoencoders demonstrate improved noise robustness and more stable performance across tasks, while significantly reducing computational and transmission costs.
\end{abstract}

\vspace{-2mm}

%
\IEEEpeerreviewmaketitle

\section{Introduction}

In recent years, foundation models (FMs) have become a dominant approach in artificial intelligence (AI), as large-scale pretraining enables broad task generalization. Specifically, a pretrained model provides learned representations that ease downstream tasks by reducing the need for extensive labeled datasets, increasing the robustness to out-of-distribution data, and reducing the complexity of downstream models. This trend was successfully adopted in Natural Language Processing (NLP), with models like GPT \cite{gpt3}, and Vision Transformers (ViT) \cite{dosovitskiy2020image}.
This paradigm has recently been adopted in telecommunications research, aiming to build a universal feature extractor through large-scale pretraining. The Large Wireless Model (LWM)~\cite{alikhani2025lwm} is a BERT-inspired architecture~\cite{devlin2018bert} proposed as a pre-trained feature extractor for wireless channels. The model structure is based on transformers and designed to allow fine-grained and global dependencies. It was pre-trained on a wide range of synthetic data through a self-supervised technique called masked channel modeling. 

The embeddings generated by LWM~\cite{alikhani2025lwm} have been successfully applied to downstream tasks, leveraging their rich representation compared to raw channel data. A key advantage highlighted is the reduced reliance on extensive labeled datasets across applications such as line-of-sight/non-line-of-sight (LoS/NLoS) classification and beam selection.
Building on these promising results, this work further explores the design space of wireless representations by providing a broader and more systematic evaluation. 
We will focus on the following aspects:

\begin{itemize}
\item \textbf{Comparison with compact representations}: We benchmark high-dimensional learned embeddings against low-dimensional autoencoder (AE)-based representations.
\item \textbf{Efficiency and overhead analysis}: We conduct a thorough analysis of data efficiency, noise robustness, and computational complexity to characterize the practical deployment trade-offs.
\item \textbf{Resource allocation as a new downstream task}: We introduce a multi-user resource allocation task to evaluate the effectiveness of different representations in a system-level setting.
\end{itemize}




\vspace{-.2cm}
\section{Background \& Preliminaries}
\vspace{-.1cm}
In this section, we present the LWM model and introduce a compact autoencoder-based baseline used for comparison in this work.
\vspace{-.1cm}
\subsection{Large Wireless Model}
\vspace{-.1cm}
The LWM model~\cite{alikhani2025lwm} uses as input the Channel State Information (CSI) $\mathbf{H} \in \mathbb{C} ^{\text{M} \times N}$ where $M$ represents the number of antennas in the Base Station (BS) and $N$, the number of subcarriers in the system. Before passing through the encoder, it is necessary to transform the CSI into patches through the following process:
split real and imaginary parts,
    flatten each component,
    divide into $P$ patches with length $L = 2 MN/P$.
Currently, the newest version of the model uses 32 as the length $L$. Those patches will be used to produce $P + 1$ embeddings with dimension $D_{\text{emb}}$ equal to 128, where the first embedding is a global summarization of the channel called CLS embedding, and the remaining ones are channel embeddings for each patch. This means that the LWM expands the dimensionality of the channel by approximately four times, which introduces additional overhead for downstream models.

During the pre-training, $p\%$ of patches are masked. Of those, 80\% are fully masked, 10\% are replaced with random vectors, and 10\% are left unchanged. The masked strategy is applied to both the real and imaginary parts to prevent information leakage. The imaginary patches selected for masking are the counterparts of the randomly selected real patches. 
The main goal of the pre-training process is to reconstruct the original patches using the embeddings generated by the LWM from the masked patches. This process is intended to enable the model to learn complex physical relations between patches and differentiate genuine channel structures from anomalies. In this work, we evaluate whether the learned representation exhibits these properties in downstream tasks.

\subsection{Compact Autoencoder Baseline}

\begin{figure}[!t]
     \centering
     \includegraphics[height=0.32\textheight]{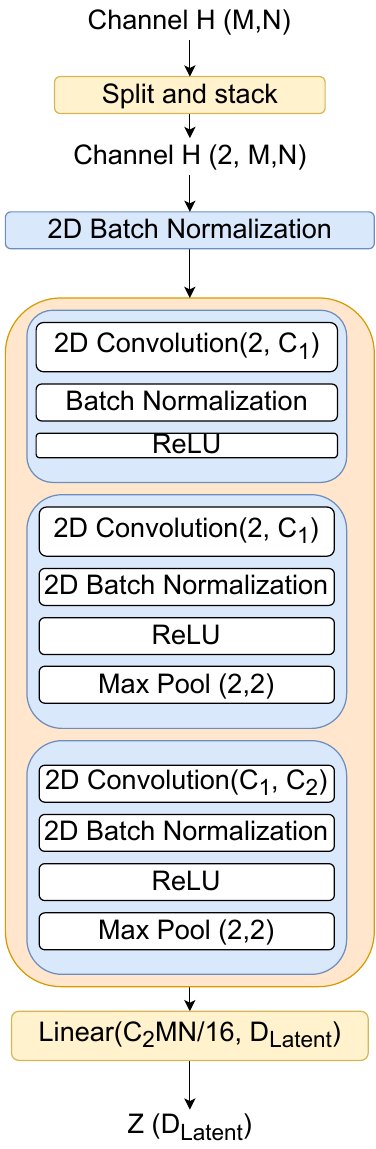}
     \vspace{-.15cm}
     \caption{Encoder architecture}
     \label{fig:csi_encoder}
     \vspace{-.55cm}
\end{figure}

As a comparable counterpart to the LWM, inspired by the deep learning-based CSI compression work initiated by CSINet~\cite{wen2018deep}, we develop a compact autoencoder-based representation, adapted and trained in this work, to serve as a benchmark and to evaluate whether the increased dimensionality of the learned embeddings provides a meaningful advantage. The baseline model is designed to be computationally efficient and lightweight.
The model consists of an encoder, shown in Fig.~\ref{fig:csi_encoder}, which maps the CSI to a compact latent representation, denoted as $\mathbf{z} \in \mathbb{R}^{D_{\text{latent}}}$, and a decoder that is used only during training to reconstruct the channel. The compression ratio of the model is defined as $D_{\text{latent}} / MN$. In this work, we adopted the ratios 1:32 (AE - 1/32) and 1:16 (AE - 1/16) for the tested autoencoders.

Specifically, the encoder processes the input through an initial 2D batch normalization layer, followed by a feature extraction module comprising three sequential convolutional blocks. These blocks apply 2D convolutions, batch normalization, and ReLU activation functions to extract spatial dependencies from the CSI matrix. To progressively downsample the spatial dimensions, the second and third blocks incorporate $2\times 2$ max pooling layers. Finally, the extracted feature maps are flattened and passed through a fully connected (linear) layer, projecting the intermediate representation of size $C_2MN/16$ into the final compressed vector of size $D_{\text{latent}}$.


The adopted training strategy consists of corrupting $\mathbf{H}$ with noise, compressing using the encoder, and decompressing while removing the noise through the decoder. The model is trained to minimize the reconstruction error between the input CSI and its reconstructed output. This encourages the model to learn a compact and noise-robust representation of the channel.

\section{Benchmarking Wireless Channel Representations via Downstream Tasks}

To evaluate the respective effectiveness of the LWM and the convolutional autoencoder as universal feature extractors, we design three downstream tasks targeting complementary capabilities: semantic understanding LoS/NLoS classification, spatial reasoning (beam selection), and resource management (power allocation).
We consider three types of representations for comparison:
\begin{itemize}
    \item \textbf{Latent embeddings} produced by the LWM transformer, including: (i) \textbf{Channel embeddings}, which capture fine-grained information from all channel patches, and (ii) \textbf{CLS embeddings}, which provide a compact global summary of the channel;
    \item \textbf{Compressed vector} generated by the autoencoder;
    \item \textbf{Raw Channel State Information (CSI)}, used as the baseline.
\end{itemize}



Finally, we analyze the different methods using three metrics: \textit{Data Efficiency} (performance vs. training samples), \textit{Noise Robustness} (performance vs. SNR), and \textit{Computational Complexity} (number of parameters, FLOPs, and inference time).

\subsection{LoS/NLoS Classification}\label{sec:task_los_nlos}

The LoS/NLoS classification is the simplest task, serving as a primary benchmark for feature quality. In wireless communications, distinguishing whether a user has a direct path to the BS is critical for beam management, handover decisions, and other tasks. 
This task evaluates whether the compressed representations preserve the physical characteristics of the propagation environment. If a model can distinguish LoS from NLoS solely from a compressed vector, it indicates that the encoder has successfully learned the underlying physics of the channel.

\subsubsection{Problem Formulation}\label{sec:task_los_nlos_problem}

The problem is modeled as a binary classification task, where the labels are generated through the DeepMIMO~\cite{Alkhateeb2019} simulation based on the selected scenario. Given an input representation $\mathbf{x}$ (derived from the channel $\mathbf{H}$), the model predicts the propagation condition $y \in \{0, 1\}$, where:
     $y = 1$ denotes a LoS condition, and $y = 0$ denotes a NLoS condition.
Since the representations have different shapes, they are flattened before entering the MLP:
\begin{itemize}
    \item \textbf{LWM}: The input $\mathbf{E} \in \mathbb{R}^{(P+1) \times D_{emb}}$ is flattened to a vector of size $(P+1)D_{emb}$.
    \item \textbf{Autoencoder}: The latent vector $\mathbf{z} \in \mathbb{R}^{D_{\text{latent}}}$ is used directly.
    \item \textbf{Raw Channel}: The complex matrix $\mathbf{H} \in \mathbb{C}^{M \times N}$ is treated as two real channels and flattened to size $2MN$.
\end{itemize}

\subsubsection{Training}\label{sec:task_los_nlos_training}

The downstream model is trained to minimize the binary cross-entropy $L = -\frac{1}{B}\sum_{i=1}^{B}\big[y_i \ln(\hat{y}_i) + (1 - y_i)\ln(1 - \hat{y}_i)\big]$,
where $y_i$ is the ground-truth label, $\hat{y}_i$ is the predicted probability and $B$ is the batch size. 






\subsubsection{Evaluation Metrics}\label{sec:task_los_nlos_evaluation}

Due to potential class imbalance (e.g., more NLoS samples in urban scenarios), we use the weighted F1-score. The weighted F1-score is defined as
$F1_{\text{weighted}} = \sum_{i \in \{0,1\}} \frac{N_i}{N} F1_i$,
where $N_i$ and $N$ denote the number of samples in class $i$ and the total number of samples, respectively, and
$F1_i = \frac{2\,\text{Precision}_i\,\text{Recall}_i}{\text{Precision}_i + \text{Recall}_i}$.

\subsection{Beam Selection}\label{sec:task_beam}

The second downstream task considers a key problem in large multiple-antenna systems: beam selection. In high-frequency multi-user systems, beamforming is essential to mitigate path loss. This task evaluates the model’s ability to capture \textbf{spatial relationships}, specifically whether the compressed representation and the learned embeddings encode sufficient geometric information to serve as effective inputs for beam selection.

\subsubsection{Problem Formulation}\label{sec:task_beam_problem}

The problem is modeled as a multi-class classification task. In this simulation, the BS is equipped with a Uniform Linear Array (ULA) serving a single-antenna user. The BS uses a pre-defined codebook structured as follows: 
\textbf{Angles:} The codebook consists of~$K$ angles spaced evenly between a field-of-view (FoV), 
$\phi_k = -\frac{\text{FoV}}{2} + k \Delta\phi$, where $\Delta\phi = \frac{\text{FoV}}{K-1}$.
\textbf{Steering vector:}~For a ULA with $M$ antennas and spacing $d$, 
$\mathbf{w}(\phi_k) = \frac{1}{\sqrt{M}}\big[1, e^{-j\frac{2\pi}{\lambda}d\sin(\phi_k)}, \dots, 
e^{-j\frac{2\pi}{\lambda}d(M-1)\sin(\phi_k)}\big]^\top$.
\textbf{Codebook matrix:}~The steering vectors are stacked to create the codebook matrix $\mathbf{F} \in \mathbb{C}^{K \times M}$
\vspace{-.2cm}
    \begin{equation}
            \mathbf{F} = \begin{bmatrix} \mathbf{w}(\phi_0)^H \\ \mathbf{w}(\phi_1)^H \\ \vdots \\ \mathbf{w}(\phi_{K-1})^H \end{bmatrix}.
    \end{equation}
    \textbf{Subcarrier averaging:}~Given $\mathbf{H} \in \mathbb{C}^{M \times N}$, the effective spatial channel is
$\bar{\mathbf{h}}_i = \frac{1}{N} \sum_{n=1}^{N} \mathbf{h}_{i,n} \in \mathbb{C}^{M \times 1}$, where $i$ represents the selected channel after averaging over the subcarriers.
\textbf{Beam Selection}:~We select the beam $k^{*}$ that maximizes the power received by the user:
\vspace{-.15cm}
    \begin{equation}
            k^{*} = \operatorname*{argmax}_{k \in \left\{ 0, \ \cdots, K-1\right\}} \left| \mathbf{w}(\phi_k)^H  \bar{\mathbf{h}}_i \right|^2
            \vspace{-.15cm}
    \end{equation}

\subsubsection{Training}\label{task_beam_training}

In this task, the model is trained to minimize the categorical cross-entropy loss over $K$ classes:
  $ L = - \sum_{k=1}^{K} y_k \ln(\hat{y}_k).$ 
It is also necessary to perform the same input adaptations as in the previous task. Finally, the task is evaluated by the same weighted F1-Score.

\subsection{Power Allocation}\label{sec:task_power}

The final downstream task evaluates the performance of the learned representation in a \textbf{resource allocation} task. Specifically, we focus on a Multi-User MIMO (MU-MIMO) system. Unlike the previous tasks, this one is a regression problem where the goal is to maximize the system's spectral efficiency under total power budget constraint.
The task verifies if the representation is enough to deal with interference patterns between users.

\subsubsection{Problem Formulation}\label{sec:task_power_problem}

Consider a system with $K$ single-antenna users served simultaneously by a BS with $M$ antennas. The received signal by the user $k$ is

\begin{equation}
    y_k = \mathbf{h}_k^H \mathbf{w}_k \sqrt{p_k} s_k + \sum_{j \neq k} \mathbf{h}_k^H \mathbf{w}_j \sqrt{p_j} s_j + n_k,
\end{equation}
where $\mathbf{h}_k \in \mathbb{C}^{M \times 1}$ is the channel vector of the $k$-th user, $\mathbf{w}_k \in \mathbb{C}^{M \times 1}$ is the beamforming vector of the $k$-th user, $s_k$ is the symbol transmitted to user $k$ with $\mathbb{E} [|s_k|^2] = 1$, $p_k$ is the allocated power for the $k$-the user, $\mathbf{h}_k^H \mathbf{w}_k \sqrt{p_k} s_k$ represents the signal for the $k$-th user, $\sum_{j \neq k} \mathbf{h}_k^H \mathbf{w}_j \sqrt{p_j} s_j $ represents the interference suffered by the $k$-th user, and $n_k$ is Additive white Gaussian noise (AWGN), $\mathcal{CN}(0, \sigma^2)$.

For the beam directions, we assume Maximum Ratio Transmission (MRT) precoding \cite{mrt1999},  $\mathbf{w}_k = \frac{\mathbf{h}_k}{\| \mathbf{h}_k \|}$. The main reason to choose this strategy is due to its low computational complexity compared to strategies that actively cancel interference, like Zero-forcing (ZF) \cite{spencer2004zf}. The resulting interference management is conducted by power allocation.
The optimization problem to be solved can be written as follows
\vspace{-.2cm}
\begin{equation}
        \begin{aligned}
            \max_{\mathbf{p}} \quad & \sum_{k=1}^{K} \log_2(1 + \text{SINR}_k) \\
            \text{s.t.} \quad & \sum_{k=1}^{K} p_k \leq P_{\text{total}}, \\
            &  (\forall k \in \{1, \dots, K\})\quad p_k \geq 0, \\[-.6cm]
            &
        \end{aligned}
        \label{eq:sum_rate_optimization}
    \end{equation}
    where $\mathbf{p} = [p_1, p_2, \dots, p_K]^T$ is the allocated power vector and $
        \text{SINR}_k = \frac{|\mathbf{h}_k^H \mathbf{w}_k|^2 p_k}{\sum_{j \ne k} |\mathbf{h}_k^H \mathbf{w}_j|^2 p_j + \sigma^2}.$

The input adaptations are the same used in the previous tasks, but accounting for the $K$.
\begin{itemize}
    \item \textbf{LWM}: Input flattened into $K$ vectors of size $(P + 1) D_{emb}$.
    \item \textbf{Autoencoder}: Latent vector $\mathbf{z}$ is transformed into a sequence of $K$ vector with size $D_{\text{latent}}$
    \item \textbf{Raw Channel}: The complex matrix $\mathbf{H}$ is treated as two real channels and flattened to $K$ vectors with size $2 M N $.
\end{itemize}

\subsubsection{Training}\label{sec:task_power_training}

Optimizing the spectral efficiency using deep learning is a difficult task due to the non-convex nature of the function. To address this problem, we choose to implement a hybrid training strategy with a fixed SNR equal to 5 dB during the training.

\paragraph{Supervised Warm-up}\label{sec:task_power_supervised}

In the first phase of the training ($60\%$ of the epochs), the model is trained using a supervised strategy to learn with a classical optimization algorithm. We generate power allocation $\mathbf{p}^{*}$ labels for a subset of training dataset using Projected Gradient Descent (PGD)~\cite{boyd2004convex}.

Specifically, the PGD algorithm iteratively maximizes the sum-rate capacity of the multi-user channel. At each iteration, it computes a gradient ascent step to increase the system throughput. To ensure the physical constraints, the algorithm then projects the resulting gradients to the simplex set $\{p_k\in[0,+\infty)^K\mid\sum_{k=1}^{K} p_k \leq P_{\text{total}}\}$).
The model is trained to minimize the Mean Squared Error (MSE) between the predicted powers $\hat{\mathbf{p}}$ and the target: $   \mathcal{L}_{\text{warmup}} = \|\hat{\mathbf{p}} - \mathbf{p}^*\|^2.$ 
This phase is very important because, due to the non-convex nature of the problem and the random initialization of the weights, the model can be stuck in a local minimum. The warm-up phase stabilizes training and provides a good initialization.

\paragraph{Unsupervised learning}\label{sec:task_power_unsupervised}

After the warm-up, the model begins an unsupervised learning phase with the objective of maximizing the spectral efficiency. Since the standard optimizers want to minimize loss, the negative SE is defined as loss: $
    \text{SE}(\hat{\mathbf{p}}) = - \frac{1}{B} \sum_{k=1}^{K} \log_2 \left( 1 + \text{SINR}_k(\hat{\mathbf{p}}) \right),$
where $B$ is the number of samples in the batch and $\text{SINR}_k(\hat{\mathbf{p}})$ is the Signal-to-Interference-plus-Noise Ratio for the $k$-th user for a given set of power $\hat{\mathbf{p}}$ for each of the $k$ users.

\paragraph{Dataset Generation: User Grouping}\label{sec:task_power_dataset}

An important aspect of MU-MIMO systems is the user selection. Randomly sampling~$K$ users could result in groups with highly correlated channels, making it impossible to manage the interference through power allocation, or very different path losses, which could result in a dominant user receiving most of the power.
To ensure a meaningful benchmark, we implemented a greedy sampling algorithm based on two criteria:
\begin{enumerate}
    \item \textbf{Spatial Correlation}: Users must have spatial correlation $\rho_{\text{min}} \leq \rho \leq \rho_{\text{max}}$ to increase the difficulty of the problem due to inter-user interference while still being able to solve it with the selected precoder.
    \item \textbf{Gain Ratio}: To minimize the near-far problem, the ratio between the weakest and strongest user should be less than $\gamma_{max}$.
\end{enumerate}

\subsubsection{Evaluation Metrics}\label{sec:task_power_evaluation}

To evaluate the performance of the downstream model based on the learned representation, we compare the spectral efficiency of predicted powers against the following baselines:
\begin{enumerate}
    \item \textbf{Equal Power Allocation (EPA)}: The naive baseline allocation, where the power is shared equally among users $p_k = P_{\text{total}} / K$.
    \item \textbf{Iterative PGD (Optimized)}: A numerical solver that runs for $i$ iterations with $r$ different random initializations.
\end{enumerate}

\section{Experimental Results}
All training and evaluation experiments were conducted on an NVIDIA L4 GPU using the Google Colab environment.
\subsection{Dataset and Downstream Models}



The dataset is generated using the DeepMIMO engine \cite{Alkhateeb2019} and \texttt{city\_18\_denver} scenario with all active BS. The simulator utilizes a ray-tracing engine to extract CSI, user coordinates, and LoS labels from scenarios based on real-world city maps. To ensure numerical stability, all channel coefficients are scaled by a normalization factor $\alpha$.

During the experiments, we considered a scenario with 32 subcarriers and a BS with 32 antennas. The dataset was divided into 3 parts: 60\% was used for training, 20\% for validation, and 20\% for testing. For the user grouping in the power allocation task, the spatial correlation is between 0.3 and 0.9, and the maximum gain ratio is 20.

To ensure a fair and controlled comparison, we employ a unified downstream architecture across all representations for each task. While a fixed architecture may favor certain representation geometries, it allows us to isolate the effect of the representations themselves and avoid confounding factors. 
For the LoS/NLoS classification, we employ a Multilayer Perceptron (MLP) that takes a flattened input, applies a linear transformation, followed by a ReLU activation, and finally a linear layer that maps the hidden representation to two output classes.
For the beam selection task, we use a fully connected network composed of three blocks, each consisting of a linear layer followed by batch normalization, ReLU activation, and dropout. The feature dimension is halved at each block, and a final linear layer maps the features to $K$ class logits.
Finally, for power allocation, we employ a 1D Residual Convolutional Neural Network (ResCNN). The input is first reshaped and passed through an initial 1D convolutional layer followed by a non-linear activation. The core of the model consists of multiple residual blocks, each comprising two 1D convolutional layers, each followed by batch normalization and a non-linear activation, with an identity residual (skip) connection to facilitate gradient flow. The network concludes with a convolutional layer that outputs one power logit per user, followed by a Softmax activation to enforce the total power constraint $P_{\text{total}}$.

\subsection{LoS/NLoS Classification Results}
\begin{figure}[t]
    \centering
    \begin{subfigure}{0.35\textwidth}
        \centering
        \includegraphics[width=\linewidth]{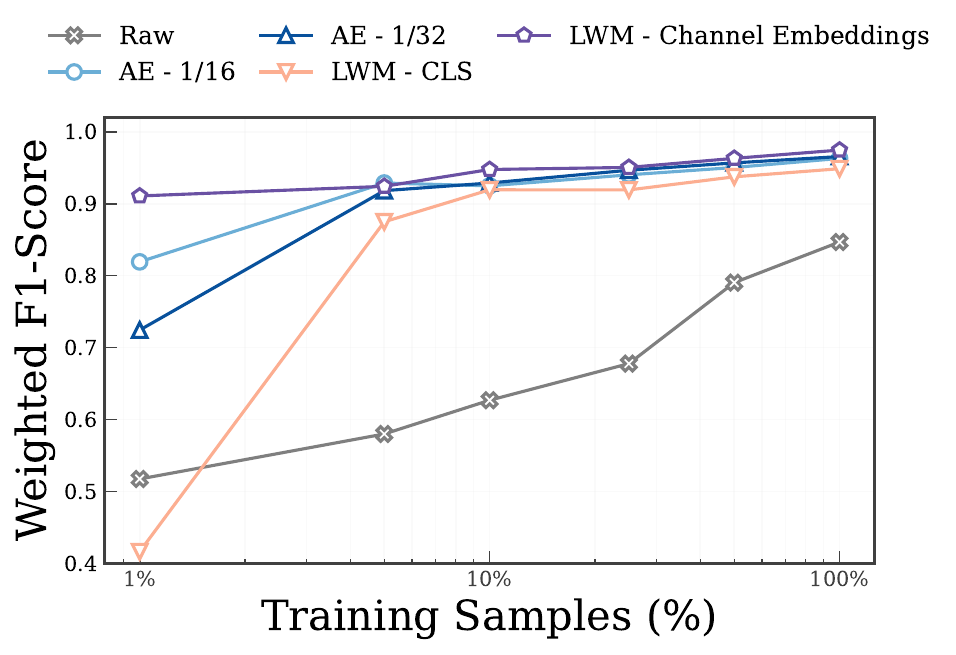}
        \vspace{-.6cm}
        \caption{Data efficiency}
        \label{fig:los_classification_data_eff}
    \end{subfigure}
    \hfill
    \begin{subfigure}{0.35\textwidth}
        \centering
        \includegraphics[width=\linewidth]{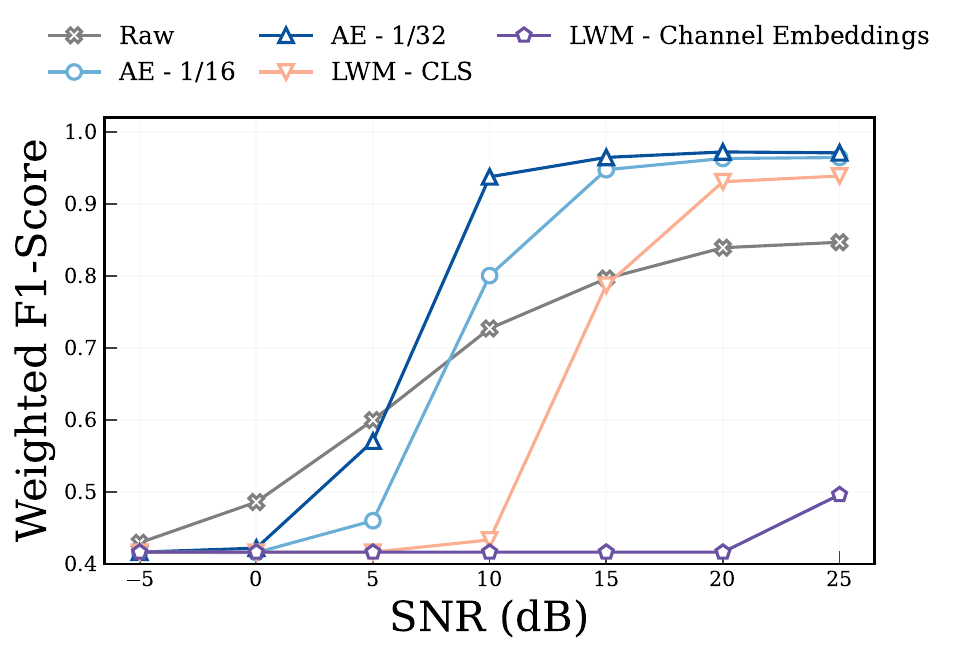}
        \vspace{-.6cm}
        \caption{Robustness to noise}
        \label{fig:los_classification_noise}
    \end{subfigure}
    \caption{LoS/NLoS classification.}
    \label{fig:two_subfigures}
    \vspace{-.35cm}
\end{figure}

Fig.~\ref{fig:los_classification_data_eff} shows that the learned representations perform well in the LoS/NLoS classification task. Channel embeddings achieve excellent performance in the few-shot regime and saturate much earlier than the raw CSI baseline; however, with sufficient data, all representations eventually converge to similar performance levels. Overall, these results highlight a clear advantage in low-data environments, where representations learned by the LWM and autoencoders significantly outperform the baseline. Although the global CLS representation may struggle in extremely low-data cases, its performance can be substantially improved through fine-tuning.



From a noise robustness perspective, all representations exhibit significant performance degradation in extremely low SNR regimes. However, the compressed vectors generated by the autoencoders demonstrate superior robustness compared to LWM embeddings (Fig.~\ref{fig:los_classification_noise}).



\begin{figure*}[htbp]
    \centering
    \includegraphics[width=0.98\linewidth]{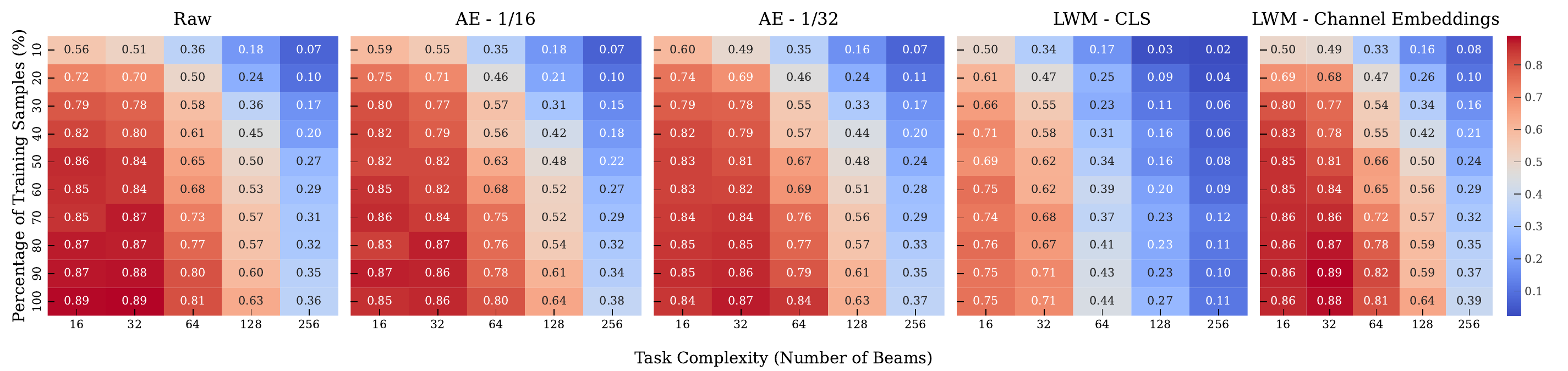}
    \caption{Data efficiency for beam selection.}
    \label{fig:beam_selection_data_eff}
\end{figure*}

\begin{figure*}[htbp]
    \centering
    \includegraphics[width=0.98\linewidth]{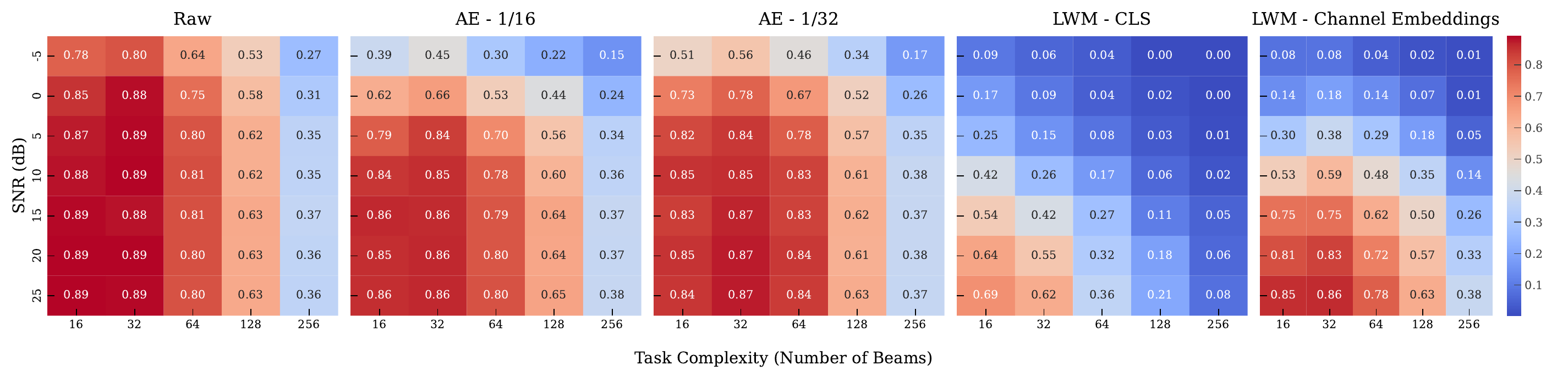}
    \caption{Robustness to noise for beam selection.}
    \label{fig:beam_selection_noise}
\end{figure*}

\subsection{Beam Selection Results}

The second set of experiments focuses on multi-class classification and best beam prediction.
In this task, we added a complexity component related to the number of available
beams. The result shows no clear advantage of the learned representations, as the compressed autoencoder vector and channel embeddings perform similarly to the baseline, while the global CLS embedding performs worse in both few-shot and high-data regimes, as illustrated in Fig.~\ref{fig:beam_selection_data_eff}. As task complexity increases, all models struggle; however, the low-dimensional compressed vector matches the performance of channel embeddings and raw CSI, making it particularly attractive for downstream tasks due to reduced computational cost and transmission bandwidth.



Finally, the learned representations show lower noise robustness compared to the baseline across all evaluated settings. In particular, LWM embeddings exhibit more pronounced performance degradation in low SNR regimes, where they are generally outperformed by both autoencoder-based representations and the raw baseline (Fig.~\ref{fig:beam_selection_noise}).

\subsection{Power Allocation Results}

\begin{figure*}[htbp]
    \centering
    \includegraphics[width=0.98\linewidth]{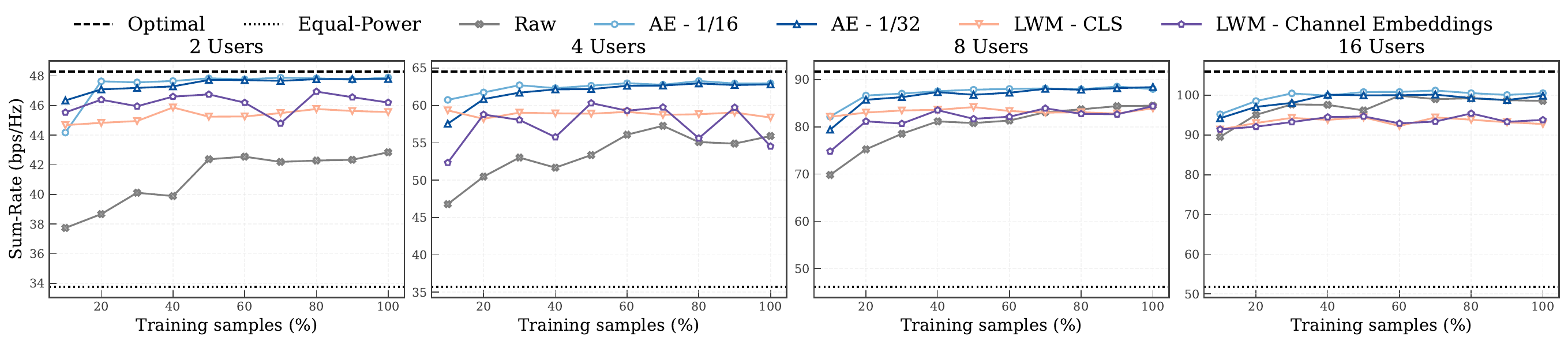}
    \caption{Data efficiency for power allocation.}
    \label{fig:power_allocation_data_eff}
\end{figure*}

In the final task, the autoencoder demonstrates higher data efficiency and improved training stability across all evaluated scenarios, outperforming both the baseline and the LWM, while achieving near-optimal performance. Also, Fig.~\ref{fig:power_allocation_data_eff} shows that as the number of users increases, the gap between representations and the optimal sum-rate also increases. This happens because the system has a greater pool of users to manage, which facilitates the power allocation.

\begin{figure*}[htbp]
    \centering
    \includegraphics[width=0.98\linewidth]{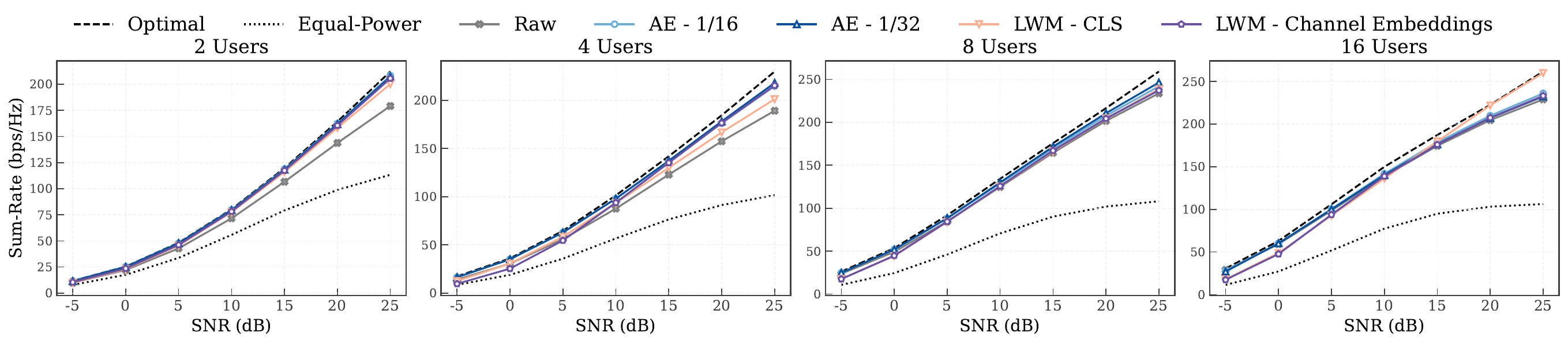}
    \caption{Robustness to noise for power allocation.}
    \label{fig:power_allocation_noise}
\end{figure*}

Fig.~\ref{fig:power_allocation_noise} has a similar robustness to noise. However, it is worth mentioning that, because of the nature of the problem, the noise also affects the capacity of the channel and the maximum achievable sum-rate. This means that while the performance of all representations decreases, the optimal solution degrades as well, partially masking the relative performance drop.

\subsection{Computational Complexity Analysis}

\begin{table}[]
    \centering
    \caption{Computational complexity for different feature extractors.}
    \begin{tabular}{cccc}
        \textbf{Model} & \textbf{Parameters (M)} & \textbf{MFLOPs} & \textbf{Inference Time (ms)} \\
        \hline
        LWM & 2.47 & 6.65 & 10.98 \\
        AE - 1/16 & 1.16 & 59.02 & 0.73 \\
        AE - 1/32 & 0.64 & 58.49 & 0.73\\
        \hline
    \end{tabular}
    
    \label{tab:computational_complexity}
\end{table}

Finally, we measured the complexity of the models used to generate the representations by calculating the number of parameters, FLOPs, and time to perform inference through a NVIDIA L4 for one sample.

In Table \ref{tab:computational_complexity}, it is clear that the LWM adds significant computational overhead during inference. Despite using fewer FLOPs to generate the embeddings, due to the memory-constrained nature of transformers, the LWM has a longer inference time and larger memory footprint. Additionally, because the foundation model has greater dimensionality, it increases the memory requirements for downstream models.

Finally, the autoencoder has one major advantage over the LWM. It is common that the UE needs to transmit the estimated CSI back to the BS. By using the autoencoder to compress the data, it generates a representation that reduces overhead during transmission and can be readily used for downstream tasks.

\section{Conclusion}

This work investigates the trade-offs associated with high-dimensional channel representations in wireless systems, motivated by recent advances in foundation models for communications. Rather than focusing on a single architecture, we provide a systematic comparison between high-dimensional learned embeddings, compressed latent representations, and raw channel data across multiple downstream tasks. Our results highlight that representation design plays a central role in determining data efficiency, robustness, and system-level overhead. In particular, while high-dimensional embeddings can offer advantages in low-data regimes for certain tasks, they introduce significant computational and transmission costs. In contrast, compact representations learned via autoencoders demonstrate strong robustness to noise and stable performance across tasks, while substantially reducing resource requirements. These findings underscore the importance of balancing representational richness with efficiency when designing learning-based wireless systems. Looking forward, this suggests a promising research direction toward architectures and training strategies that jointly optimize performance, robustness, and deployment efficiency in realistic communication environments.






%
\bibliographystyle{IEEEtran}
\bibliography{collection}

\end{document}